\documentstyle[psfig]{mn}

\newif\ifAMStwofonts

\ifoldfss
  \ifCUPmtlplainloaded \else
    \NewTextAlphabet{textbfit} {cmbxti10} {}
    \NewTextAlphabet{textbfss} {cmssbx10} {}
    \NewMathAlphabet{mathbfit} {cmbxti10} {} 
    \NewMathAlphabet{mathbfss} {cmssbx10} {} 
  \fi
  \ifAMStwofonts
    \ifCUPmtlplainloaded \else
      \NewSymbolFont{upmath} {eurm10}
      \NewSymbolFont{AMSa} {msam10}
      \NewMathSymbol{\upi}     {0}{upmath}{19}
      \NewMathSymbol{\umu}     {0}{upmath}{16}
      \NewMathSymbol{\upartial}{0}{upmath}{40}
      \NewMathSymbol{\leqslant}{3}{AMSa}{36}
      \NewMathSymbol{\geqslant}{3}{AMSa}{3E}

      \let\leq=\leqslant \let\le=\leqslant
      \let\geq=\geqslant 
    \fi
  \fi
\fi 

\ifnfssone
  \newmathalphabet{\mathit}
  \addtoversion{normal}{\mathit}{cmr}{m}{it}
  \addtoversion{bold}{\mathit}{cmr}{bx}{it}
  \newmathalphabet{\mathbfit} 
  \addtoversion{normal}{\mathbfit}{cmr}{bx}{it}
  \addtoversion{bold}{\mathbfit}{cmr}{bx}{it}
  \newmathalphabet{\mathbfss} 
  \addtoversion{normal}{\mathbfss}{cmss}{bx}{n}
  \addtoversion{bold}{\mathbfss}{cmss}{bx}{n}
  \ifAMStwofonts
    \ifCUPmtlplainloaded \else
      \UseAMStwoboldmath
      \makeatletter
      \new@mathgroup\upmath@group
      \define@mathgroup\mv@normal\upmath@group{eur}{m}{n}
      \define@mathgroup\mv@bold\upmath@group{eur}{b}{n}
      \edef\UPM{\hexnumber\upmath@group}
      \new@mathgroup\amsa@group
      \define@mathgroup\mv@normal\amsa@group{msa}{m}{n}
      \define@mathgroup\mv@bold\amsa@group{msa}{m}{n}
      \edef\AMSa{\hexnumber\amsa@group}
      \makeatother
      \mathchardef\upi="0\UPM19
      \mathchardef\umu="0\UPM16
      \mathchardef\upartial="0\UPM40
      \mathchardef\leqslant="3\AMSa36
      \mathchardef\geqslant="3\AMSa3E

      \let\leq=\leqslant \let\le=\leqslant
      \let\geq=\geqslant 
    \fi
  \fi
\fi 

\ifnfsstwo
  \DeclareMathAlphabet{\mathbfit}{OT1}{cmr}{bx}{it}
  \SetMathAlphabet\mathbfit{bold}{OT1}{cmr}{bx}{it}
  \DeclareMathAlphabet{\mathbfss}{OT1}{cmss}{bx}{n}
  \SetMathAlphabet\mathbfss{bold}{OT1}{cmss}{bx}{n}
  \ifAMStwofonts
    \ifCUPmtlplainloaded \else
      \DeclareSymbolFont{UPM}{U}{eur}{m}{n}
      \SetSymbolFont{UPM}{bold}{U}{eur}{b}{n}
      \DeclareSymbolFont{AMSa}{U}{msa}{m}{n}
      \DeclareMathSymbol{\upi}{0}{UPM}{"19}
      \DeclareMathSymbol{\umu}{0}{UPM}{"16}
      \DeclareMathSymbol{\upartial}{0}{UPM}{"40}
      \DeclareMathSymbol{\leqslant}{3}{AMSa}{"36}
      \DeclareMathSymbol{\geqslant}{3}{AMSa}{"3E}

      \let\leq=\leqslant \let\le=\leqslant
      \let\geq=\geqslant 
    \fi
  \fi
\fi 

\ifCUPmtlplainloaded \else
  \ifAMStwofonts \else 
    \def\upi{\pi}
    \def\umu{\mu}
    \def\upartial{\partial}
  \fi
\fi

\title{Restrictions to the galaxy evolutionary models from the Hawaiian Deep
Fields  SSA13 and SSA22}
\author[J. A. L. Aguerri \& I. Trujillo]
       {J. A. L. Aguerri$^{1}$ \& I. Trujillo$^{2}$\\
        1.-Astronomisches Institut der Universit\"{a}t Basel. Venusstrasse 7. CH-4102 Binningen. Switzerland. \\
	2.-Instituto de Astrof\'{\i}sica de Canarias. E38200 La Laguna (Tenerife). Spain.}
\date{Accepted 0000 December 00.
      Received 0000 December 00;
      in original form 0000 October 00}

\pagerange{\pageref{firstpage}--\pageref{lastpage}}
\pubyear{1994}

\begin{document}

\maketitle

\label{firstpage}

\begin{abstract} Quantitative structural analysis of the galaxies present in
the Hawaiian Deep Fields SSA13 and SSA22 is reported. The structural parameters
of the galaxies have been obtained  automatically  by fitting a two--component
model (S\'ersic $r^{1/n}$ bulge and exponential disc) to the surface brightness
of the galaxies. The galaxies were classified on the basis of the
bulge-to-total luminosity ratio ($B/T$). The magnitude selection criteria and
the reliability of our method have been checked by using Monte Carlo
simulations.  A complete sample of objects up to redshift 0.8 has been
achieved. Spheroidal objects (E/S0) represent $\approx 33\%$ and spirals 
$\approx 41\%$  of the total number of galaxies, while mergers and unclassified
objects represent  $\approx 26\%$. We have computed the comoving space density
of the different kinds of objects. In an Einstein--de Sitter universe a
decrease in the comoving density of E/S0 galaxies is observed as redshift
increases (a $\approx30\%$ less at z=0.8), while for spiral galaxies a
relatively quiet evolution is reported. The framework of hierarchical
clustering evolution models of galaxies seems to be the most appropriate to
explain our results. \end{abstract}

\begin{keywords}
galaxies: distances and
redshift---galaxies: evolution---galaxies: photometry---galaxies: fundamental
parameters
\end{keywords}

\section{Introduction}

Achieving a good galactic evolutionary model  is one of the challenges of
present astronomy. The high quality of the Hubble Space Telescope (HST) data
allows astronomers to study  the evolution of galaxy morphology  over a
significant fraction of the age of the Universe,  restricting  the two main
present theoretical frameworks of galaxy evolution: the monolithic collapse and
the hierarchical clustering models.

The simplest models of galaxy evolution predict that massive elliptical
galaxies  formed at high redshift in a rapid collapse with a single burst of
star formation (Eggen et al 1962; Larson 1975). Against this scenario, the
hierarchical clustering models predict that the most massive objects form at
late times via the merging of smaller subunits (White \& Rees 1978; Kauffmann et
al 1993). Each model has very different observational implications  (e.g.
Brinchmann et al 1998; Schade et al 1999; Fried et al 2001). Observational
evidence has been found for both scenarios (see Schade et al 1999 and references
therein), so that the dominant mechanism of galaxy evolution remains an open
question.

Many attempts have been made to classify galaxies on HST deep images. Two
families of methods are currently used: visual and automated classifications.
Among visual classifications we  mention analysis done by van der Bergh et al
(1996, 2000) in the range 21$<$I$_{814}$$<$25 at the Hubble Deep Field (HDF). They
found that up to 30$\%$ of the galaxies were ellipticals, the remainder divided
into 31$\%$ spirals and 39$\%$ unclassified. Possible differences in the
morphologies of galaxies at high redshifts point to different the environmental
conditions of these galaxies relative to the local ones. In particular, the
merger rate could be very different.  Le Fevre et al (1999) have found that the
rate of mergers and interaction  grows strongly with the redshift. Quantitative
classification systems based on the study of the central concentration and
asymmetry of the galaxian light (Abraham et al 1996) also obtained a high
fraction of irregular and peculiar galaxies at high redshifts, finding only a
20$\%$ of elliptical fraction.

Most sophisticated classification systems based on the decomposition of the
surface brightness profiles of  galaxies into their structural components
(bulge and disc mainly) have been applied during the last few years. This
technique is  used extensively for local galaxies (see Prieto et al 2001 and
references therein) but the lower resolution  at high redshift makes its
application there more difficult . This quantitative classification method has
the advantage that it gives information about each  component of  galaxies.
This means that we can follow the evolution of different components (bulge and
disc in spirals) as a function of  redshift.  Usually, it is assumed that the
same type of profiles which fit the light distribution of local galaxies also
describe the light distribution of galaxies at higher redshift. Typically,
S\'ersic r$^{1/n}$ profiles are fitted to the surface brightness profiles of bulges and
elliptical galaxies, and exponential profiles to the discs of Spiral galaxies
(Marleau \& Simard 1998, Schade et al 1996, 1999). 

Using this decomposition technique on the HDF, Marleau \& Simard (1998) found a
substantially different result from those obtained by visual classifications. 
They found that only 8\% (versus 30\% for visual classifications) of the galaxy
population down to $I_{814}(AB)=26$ are spheroidal systems. Although
quantitative methods have clear advantages over visual methods, they are not
free from significant bias which affect  the reliability of the physical
properties obtained. In order to understand the big discrepancy pointing out in
the previous analysis, it is crucial to remove the bias which are present in
quantitative analysis methods.

In this paper, we examine the structural properties of the galaxies in two
Hawaiian Deep Fields (SSA13 and SSA22) imaged by HST. Each of these fields is
composed by 3 HST/WFPC2 fields. All the galaxies studied in these fields have
spectroscopic redshifts, avoiding a strong source of uncertainty at the
distance determination. Previous classification schemes of high redshift
galaxies from HST images are  compared with our results. In particular, we
focus our attention on evaluating the number of spheroidal systems in  field
galaxies and on constraining the two main theoretical frameworks of galaxy
evolution.

The structure of this paper is as follow: section 2 describes the
characteristics of the observed fields. The structural decomposition method is
presented in section 3. In section 4 we discuss the completeness of the
sample and we summarize our conclusions and discuss their implications for galaxy
evolution in sections 5 and 6.

\section[]{The sample of galaxies}

The sample consists on all objects with K$<$20, I$<$22.5 (Kron-Cousins) and
B$<$24.5 in two areas surrounding the Hawaii deep survey fields SSA 13 and SSA
22 (Cowie et al 1994; Songaila et al 1994). Hereafter, the I magnitude will be
given in the same system as in Cowie et al (1996). Nearly all objects included
in those fields have measured spectra with the LRIS spectrograph on Keck (Cowie
et al 1996) \footnote{See the discussion about the different magnitudes and
transformations in Cowie et al (1995)}. The fields were imaged during  2000 s
with the WFPC2 at HST in the I$_{814}$ bandpass. The total analyzed sky area 
was 28 arcmin$^{2}$. The analyzed objects lie in the redshift interval
$[0.1,1.3]$,  mainly concentrated  around $z=0.5$ (see Cowie et al 1996).

We used the SExtractor galaxy photometry package (Bertin \& Arnouts 1996,
version 2.1.4) for the extraction of the objects from the public released HST
images. This package is optimized to detect and measure sources from
astronomical images. The detection was run using the same parameters as in
Marleau \& Simard (1998). In particular, we used a detection threshold of
1.5 $\sigma$, where $\sigma$ is the standard deviation of the sky background of
the images. Another important parameter is the deblending parameter. SExtractor 
deblends objects using multiple flux thresholding. The SExtractor deblending 
parameter sets the minimum fraction of the total flux a branch must contain to 
be considered a separate object. We have use the same value that 
Marleau \& Simard (1998), which is 0.001.

In order to obtain a bulge+disc decomposition of the objects, we fit
ellipses to their isoluminosity contours down to 1.5 $\sigma$ using the task
ELLIPSE from IRAF. The surface brightness and ellipticity profiles obtained are
used to recover the structural parameters of the galaxies.  

\begin{figure}
 \centerline{\psfig{figure=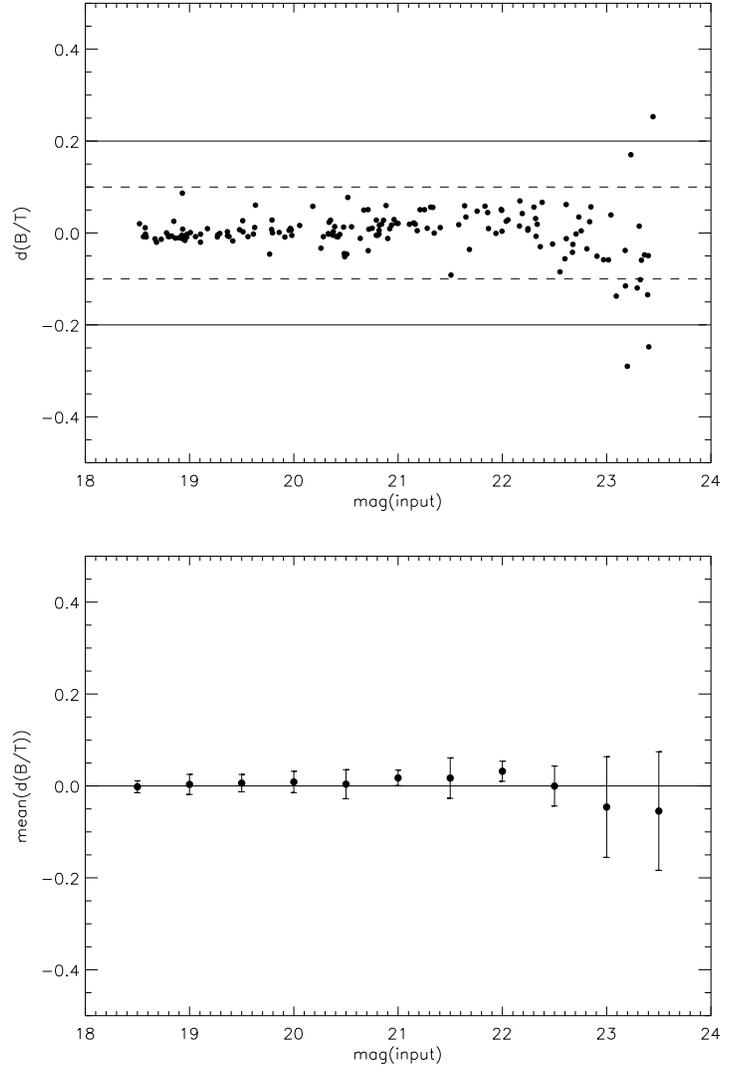,angle=0,width=10cm}}
  \caption{(Top) $d(B/T)$=$B/T$(measured)--$B/T$(input) as a function of the
input magnitude. (Bottom) Mean d(B/T) vs. input magnitude with 1 $\sigma$ error bars.}
\label{Fig:NparNprof}
  \end{figure}

\section{The galaxy classification procedure}

The classification technique is based on the decomposition of the surface
brightness profiles of the galaxies in bulge and disc components. The fitting
algorithm is discussed extensively  in Trujillo et al (2001b). Here we explain
the main points of the routine.  

The  final surface brightness distributions resulting from the convolution
between the PSF and our 2D (i.e. elliptical) model surface brightness
distributions are dependent on the  intrinsic ellipticity of the original
source - as is the case with real data.   A key problem remains, which is what
value of the ellipticity is choosen to represent the ellipticity of the
model.   The ellipticity of the isophotes are reduced by  seeing.  This
reduction depends on the radial distance of the isophote to the center of the
model, the size of the seeing, and the values of the model parameters.
Consequently, to evaluate the intrinsic ellipticity of a model it is often
insufficient to simply measure the ellipticity at one given radial distance
(e.g. 2 effective radii).   To illustrate this, the observed ellipticity at 2
r$_e$ on galaxies which have an effective radius of similar size to the FWHM
(these galaxies are common at high redshift) is 30\% less than the true 
ellipticity for galaxies with an exponential profile (n=1), and 45\% less for
galaxies with a de Vaucouleurs profile (n=4). The use of models with
underestimated ellipticity affects the evaluation of the other model
parameters, biasing the results. One result of this bias is the estimation of
smaller values of index n. This bias increases as  the value of n increases
(Trujillo et al. 2001a,c).

Consequently, the determination of the intrinsic ellipticity of the source and
the fitting process to determine the structural parameters should be done in 
tandem (i.e. using an iterative and self-consistent routine) and not as two
separate tasks.  To do this we simultaneously fit both the observed surface 
brightness and ellipticity profiles using convolved profiles for each (see how
the algorithm works in Figure 6 from Trujillo et al. 2001b). 

Our 2D fitted galaxy model has two components: a bulge and a disc.  The 2D
bulge component is a pure S\'ersic (1968) profile of the form\footnote{The
surface brightness distribution are explicitily written on elliptical
coordinates ($\xi,\theta$) (Trujillo et al. 2001a).}:

\begin{equation}
I(\xi)=I_{e}10^{-b_n[(\xi/r_{e})^{1/n}-1]}
\end{equation}
 
where $I_{e}$ is the effective intensity, $r_{e}$ is the bulge effective radius
 and $b_{n}=0.868n-0.142$ (Caon et al 1993). The disc component is an
 exponential profile given by:

\begin{equation}
I(\xi)=I_{o}e^{\xi/h}
\end{equation}

where $I_{o}$ is the central intensity and $h$ is the exponential disc
scale-length. The set of free parameters is completed with the ellipticities of
the bulge $\epsilon_{b}$ and the disc $\epsilon_{d}$. The bulge and disc
profiles were convolved with the instrumental PSF of the HST obtained from 
stellar profiles located on the images. Special attention was paid to  this
convolution. The real PSFs were fitted by Moffat functions and the convolutions
were developed analytically on real space. Also, to avoid the problem of the
undersampling of the PSF we average different stellar profiles obtaining a
composed median PSF. To this median profile we fit our analytical PSF. We have
estimated a $\sim$5\% uncertainty  in the estimation of the FWHM due to changes
from one WFPC2 position to another. This uncertainty implies an error on the
parameters estimation less than 10\%. 

\begin{figure}
 \centerline{\psfig{figure=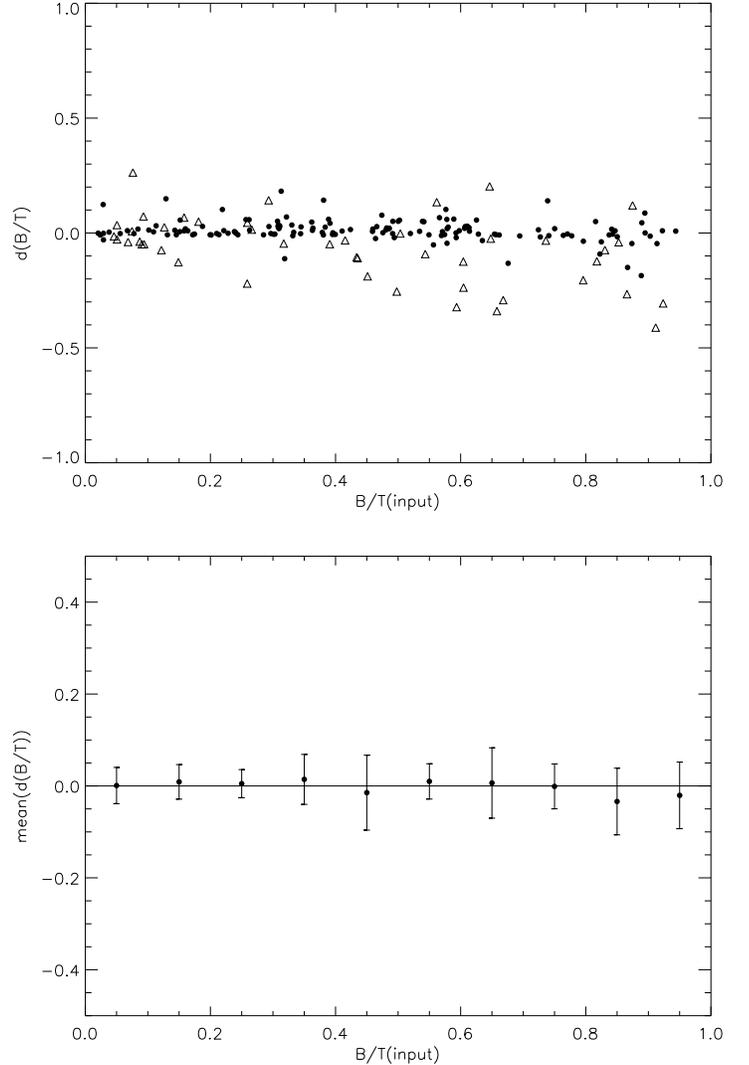,angle=0,width=10cm}}
  
  \caption{(Top) The difference, d(B/T), between measured and input B/T versus
B/T(input) for two different magnitude intervals: $I\leq 23$ (solid circles)
and $I$>$23$ (triangles). (Bottom) Mean d(B/T) versus B/T(input)  with 1 $\sigma$
error bars.}

\label{Fig:NparNprof}
  \end{figure} 

A Levenberg-Marquardt non--linear fitting algorithm (Press et al 1992) was used
to determine the free parameters set which minimizes $\chi^{2}$. Extensive
Monte-Carlo simulations were done in order to check the reliability of the
recovered parameters (see Section 4). The surface brightness profiles and 
ellipticity profiles of each galaxy were fitted at the same time. Each galaxy
was fitted by a single S\'ersic profile and a S\'ersic + exponential profile.

Following previous studies (e.g. Marleau \& Simard 1998),  galaxy
classification was based on the bulge to total luminosity ratio, $B/T$. We
consider as ``ellipticals''  those objects with $B/T>0.6$, in which case a
better fit can be obtained with  only one component. The parameters of these
objects were taken from the pure S\'ersic fitting. Galaxies with $B/T$ between
0.5--0.6 were classified as S0. Finally, objects with $B/T<0.5$ were classified
as ``spirals''. We consider as ``spheroidal'' galaxies those with $B/T>$0.5 as
Marleau \& Simard (1998). The discrimination between the different types of 
galaxies was made following  the values of $B/T$ given by Simien \& de
Vaucouleurs (1986). Quantitative selected ellipticals can be contaminated by
galaxies such as compact narrow emission--line objects. These galaxies exhibit
high bulge fractions even though they are not real ellipticals. These objects
can be $\sim$15\% of the elliptical sample (Im et al. 2001) and their presence must
be taken into account at estimating the uncertainty on the ellipticals comoving
density parameters\footnote{As a matter of caution we also must regard that
these ``interlopers'' are basically placed at high redshift (z$>$0.8) or they are
faint galaxies $M_B<$-18 (see Im et al. 2001 their Figure 17). For that reason,
most of these galaxies are expected to be out of our studied sample.}.

Galaxies selected  using only  the $B/T>$0.5 criteria may not all be E/S0s, but
could include later galaxy types. To quantify this bias we use the analysis
performed by Im et al. (2001) for a local galaxy sample (Frei et al. 1999).
Galaxies with $T\leq$0 (i.e. E/S0s) represent  76\% of the local sample
selected using $B/T>$0.5. So, a contamination of $\sim$ 25\% can be expected
in the objects that we are labeling as E/S0s at high redshift. However, the
contamination for objects with $T\leq$0 in  the objects named ``spirals'' (i.e.
$B/T<$0.5) is just 8\%. Some methods have been identified  to remove the
bias in the E/S0s selected sample with the use of red colors selected galaxies
or the use  of low asymmetry objects. However, the first option clearly biases
the sample to objects that have a quiet evolution (and what we want is
precisely study this hypothesis) and the second has been shown to be
inappropriate in objects at high redshift (i.e. low S/N as our objects have) by
Conselice,  Bershady \& Jangren (2000). Despite the known morphological type
biases, due to the above reasons we have chosen to maintain the B/T selection
criteria as the sole morphological selection criteria.

For the ellipticals, we have also imposed a restriction based on absolute
magnitude. By doing this, we have classified a galaxy as ``dwarf'' when 
$M_{B}\geq -17.0$. The absolute magnitudes were obtained after applying the
K-correction prescription of  Poggianti (1997) assuming (hereafter) a cosmology with 
H$_0$=75 km s$^{-1}$ Mpc$^{-1}$, q$_0$=0.5, $\Omega_{m}=1.0$ and
$\Omega_{\Lambda}$=0. 

Once the automated classification is done,  a visual inspection was also made
for each object. Some objects are not fitted well by either a pure S\'ersic
profile or a bulge+disc profile. They were classified as ``irregular''
galaxies. Those with evidence of mergers (close companions and irregular
shapes) were catalogued as ``mergers''. We  also had 4 objects whose
best fit is achieved by a pure S\'ersic profile with $n\approx0.5$. It is
important to note that the luminosity density of a S\'ersic profile with n$<$
0.5 has a depression in its central part representing an unlikely physical
situation (Trujillo et al 2001a). Marleau \& Simard (1998) also obtained some
objects of this class on the HDF images. The visual morphological shape of
these objects is peculiar  appearing elongated. Marleau \& Simard
(1998) claimed that this kind of objects could be remnants of mergers or close
tidal disruptions. We have included them  into the merger category. 

\begin{figure}
 \centerline{\psfig{figure=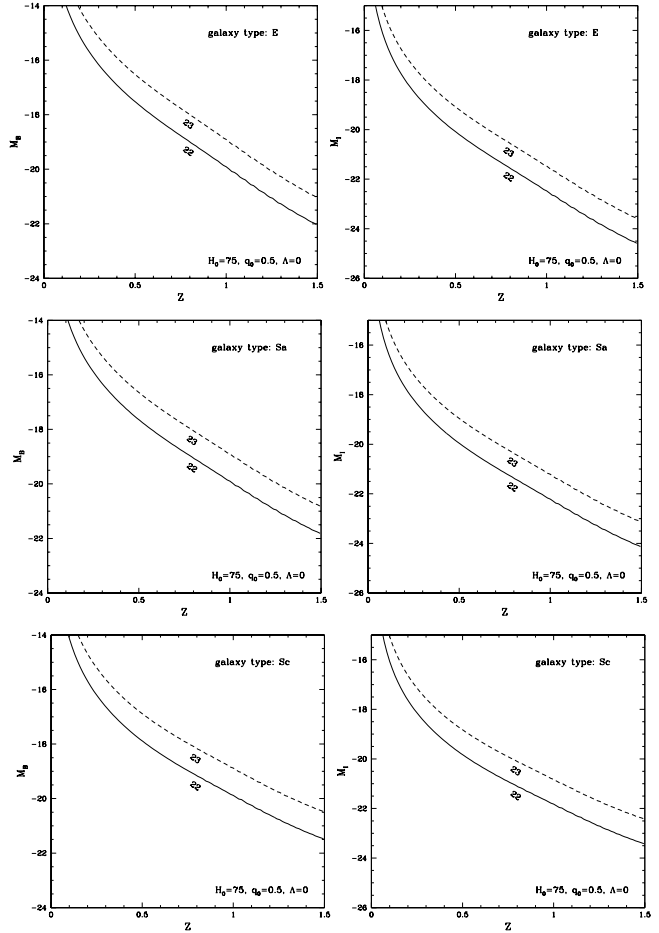,angle=0,width=10cm}}
  
  \caption{The complete magnitude as a function of the redshift for objects with
apparent magnitudes: $I=22$ (full line) and $I=23$ (dashed line). Three
different kind of objects are represented: ellipticals (top), Sa (middle) and Sc
(bottom). See text for details.}

\label{Fig:NparNprof}
  \end{figure} 

\section{The completeness of the sample}

Since selection effects can mimic  evolutionary changes in high redshift
objects, it is necessary to achieve a complete unbiased sample of objects. The
determination of the completeness of the sample is done in two steps.  First,
we determine the faintest apparent magnitude down to which the recovered
parameters are reliable. In particular, we will focus on the $B/T$ ratio
because it is the parameter used for the classification of the galaxies. We
evaluated this limiting magnitude by Monte--Carlo simulations of artificial
galaxies with similar magnitudes and structural parameters as the real objects.
Once this magnitude is obtained, the second step for the completeness of the
sample consists in determining how bright (i.e. the absolute magnitude)  a
galaxy has to be  in order to be observed in our whole  redshift interval. The
limiting absolute magnitude was obtained  by using typical spectra from every
type of objects, which allows us to verify that we are studying the same kind
of objects in all redshift intervals. Unfortunately, most   previous studies of
the structural properties of  high redshift samples  do not determine   their
limiting absolute magnitudes. This type of samples are obtained with only an
apparent limiting magnitude, which biases the sample  to the brightest objects
at high redshifts.  For this reason, it is   necessary to use a sample cut by
absolute magnitude.

\begin{figure}
 \centerline{\psfig{figure=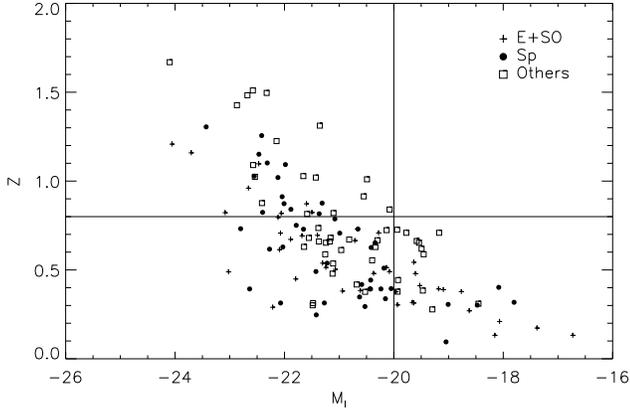,angle=0,width=9cm}}
  \caption{The $M_{I} - z$ diagram for the galaxies detected at the SSA13 and SSA22 fields. The absolute magnitudes have been computed for a $H_{o}=75 km s^{-1} Mpc^{-1}$, $\Omega_{m}=1$  and $q_{o}=0.5$ cosmology.}
\label{Fig:NparNprof}
  \end{figure} 

\subsection{Monte-Carlo simulations}

We  performed Monte--Carlo simulations to test the reliability of our
method. First, we tested the ability to recover parameters from bulge--only
(i.e. purely elliptical) structures, and second we explored the possibility of
carrying out accurate bulge+disc decompositions. In both cases we created 150
artificial galaxies with structural parameters randomly distributed in the
following ranges:

\begin{itemize}

\item  bulge--only structures: 19$\leq$I$\leq$23, 0.05$''$$\leq$$r_{e}
\leq$0.6$''$, 0.5$\leq n \leq 4$, and $0\leq \epsilon \leq 0.6$ (the lower limit
on $n$ is due to the physical restrictions pointed out in Trujillo et al
2001a).

\item  \ bulge+disc \ structures: 18.5$\leq$I$\leq$23.5, 0.05$''$$\leq$$r_{e}
\leq$0.6$''$, 0.5$\leq n \leq 4$, and $0\leq \epsilon_{b} \leq 0.4$, 0.2''$\leq
h \leq 1.5 ''$, 0$\leq B/T \leq 1$, and $0\leq \epsilon_{d} \leq 0.6$.
\end{itemize}

The artificial galaxies were created by using the IRAF task MKOBJECT. We
support as an input to this task the surface brightness distribution coming
from our detailed convolution between the PSF and the original model.  To
simulate the real conditions of our observations, we added a background sky
image (free of sources) taken from a piece of the real image; the dispersion in
the sky determination was 0.1 \%. The PSF FWHM in the simulation was set at
0.2$''$  and assumed  known exactly. The pixel scale of the simulation was
0.1$''$, as is the real WFPC2 pixel size. The same procedure was used to
process both the simulated and the actual data.

\begin{figure}
 \centerline{\psfig{figure=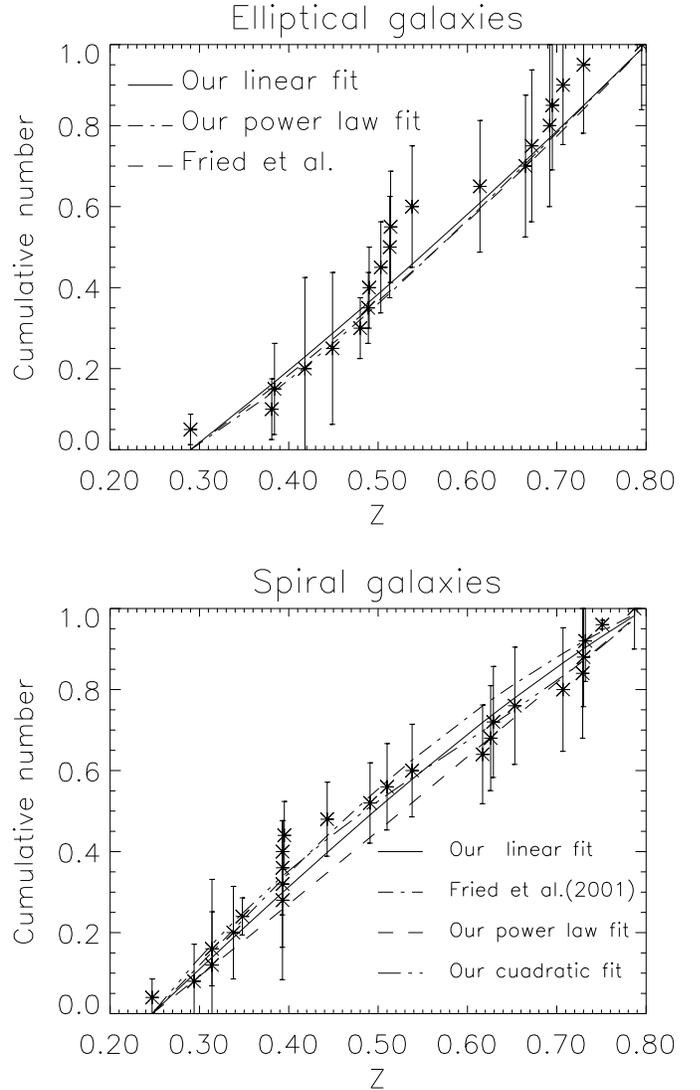,angle=0,width=10cm}}
  \caption{The cumulative number distribution function of E/S0s (top)
and spirals (bottom) as a function of the redshift. The best fits derived 
from linear and quadratic comoving densities are overplotted. It is also showed the cumulative distributions derived from Fried et al. (2001). See text for
more details.}
\label{Fig:NparNprof}
  \end{figure}

Figure 1 shows $d(B/T)$=$B/T$(measured)--$B/T$(input) as a function of the
input magnitude.  For galaxies brighter than $I=23$ magnitude $d(B/T)$ is less
than 0.1. This is a very accurate determination of this parameter. Figure 2
shows $d(B/T)$ as a  function of the $B/T(input)$. Objects with $I \geq 23$
(triangles) have bigger  dispersion of $d(B/T)$ than objects with $I \leq 23$
(points). From our simulations it follows that $I=23$ is the limiting magnitude
for   reliable recover of the $B/T$ parameter. The limiting magnitude for the
rest of structural parameters will be studied in a forth--coming paper. To our
limiting apparent magnitude, the sample of galaxies is reduced to 120 galaxies.
According to their $B/T$ ratio, absolute magnitude  and visual inspection (see
Section 3), they were classified as: ellipticals (26), dwarfs (6), S0 (9),
irregulars (20), mergers (17) and spirals (42). This left us with $\sim34\%$
spheroidal galaxies (E+Dwarfs+S0), $\sim35\%$ spirals and $\sim31\%$ of
unclassified objects. Tables 1 and 2 show the B/T ratios (column 4) and
$M_{I}$  (column 3) for the 120 galaxies with $I \leq 23.0$ for the SSA13 and
SSA22, respectively. These tables also show the identification number (column
1) and the redshift of the objects (column 2) given by Cowie et al (1996).
Galaxies classified as  ellipticals and dwarfs have $B/T=1.0$, those classified
as irregular of mergers have been marked with $B/T=-1.0$.

\begin{table}
 \centering
  \caption{Galaxies from SSA13 with $I \leq 23.0$.}
  \begin{tabular}{cccccccc}
\hline
ID  &    Z    & $M_{I}$  & B/T     &ID     &	 Z   & $M_{I}$  & B/T	 \\
\hline				   				      
5   &  0.612  &  -20.96  & -1.00   &  67   &  0.270  &  -18.62  &  1.00  \\
10  &  0.554  &  -20.39  & -1.00   &  69   &  0.317  &  -19.66  &  1.00  \\
11  &  1.225  &  -22.14  & -1.00   &  70   &  0.314  &  -21.27  &  0.32  \\
12  &  0.489  &  -23.02  &  0.54   &  71   &  0.210  &  -18.06  &  1.00  \\
14  &  0.667  &  -20.29  & -1.00   &  72   &  0.876  &  -21.31  &  0.05  \\
16  &  1.614  &  -28.14  &  1.00   &  75   &  0.818  &  -22.06  &  1.00  \\
18  &  0.491  &  -21.42  &  0.20   &  78   &  0.490  &  -20.08  &  1.00  \\
19  &  0.393  &  -20.42  &  0.03   &  87   &  1.427  &  -22.86  & -1.00   \\
20  &  1.028  &  -21.65  & -1.00   & 100   &  0.377  &  -20.52  & -1.00  \\
21  &  0.443  &  -20.42  &  0.10   & 101   &  1.256  &  -22.41  &  0.32  \\
25  &  0.730  &  -20.65  &  0.03   & 103   &  0.629  &  -22.02  &  0.31   \\
28  &  0.736  &  -21.38  & -1.00   & 105   &  0.395  &  -20.04  &  0.40  \\
31  &  1.090  &  -22.57  & -1.00   & 107   &  0.314  &  -19.63  &  1.00  \\
32  &  0.278  &  -19.29  & -1.00   & 108   &  0.680  &  -21.55  & -1.00  \\
36  &  0.338  &  -20.15  &  0.25   & 109   &  0.393  &  -20.43  &  0.30  \\
37  &  1.020  &  -21.42  & -1.00   & 110   &  0.660  &  -21.17  & -1.00  \\
38  &  0.393  &  -20.23  &  0.28   & 111   &  0.729  &  -21.65  &  0.07  \\
39  &  0.449  &  -21.79  &  1.00   & 113   &  0.629  &  -20.33  & -1.00  \\
41  &  0.480  &  -20.37  &  0.57   & 114   &  0.660  &  -21.37  & -1.00  \\
43  &  1.305  &  -23.43  &  0.34   & 115   &  0.389  &  -19.09  &  0.52  \\
46  &  0.820  &  -21.10  & -1.00   & 116   &  0.630  &  -21.64  & -1.00  \\
47  &  0.732  &  -22.80  &  0.44   & 120   &  0.841  &  -21.88  &  0.02  \\
52  &  0.914  &  -20.54  & -1.00   & 122   &  0.503  &  -21.06  &  1.00 \\
55  &  1.028  &  -22.55  &  0.22   & 124   &  0.393  &  -22.63  &  0.16 \\
59  &  1.483  &  -22.67  & -1.00   & 127   &  0.393  &  -19.18  &  1.00 \\
61  &  0.310  &  -18.45  & -1.00   & 155   &  0.730  &  -20.66  &  0.57 \\
62  &  0.314  &  -22.07  &  0.43   & 171   &  0.726  &  -19.94  & -1.00  \\
64  &  0.681  &  -21.15  & -1.00   & 174   &  0.479  &  -19.60  &  1.00  \\
\hline
\end{tabular}
\end{table}

\begin{table}
 \centering
 \begin{minipage}{140mm}
  \caption{Galaxies from SSA22 with $I \leq 23.0$.}
  \begin{tabular}{cccccccc}
\hline
   ID     &    Z        & $M_{I}$ & B/T & ID     &    Z        & $M_{I}$ & B/T\\
\hline
9   &  1.093  &  -21.98  &  0.04 & 80   &  1.669  &  -24.09  & -1.00   \\
11  &  0.626  &  -20.41  &  0.19 & 81   &  0.384  &  -19.47  & -1.00   \\
13  &  0.653  &  -20.34  &  0.34 & 82   &  0.384  &  -20.61  &  1.00   \\
14  &  0.538  &  -21.30  &  1.00 & 83   &  0.510  &  -20.18  &  0.12   \\
19  &  0.294  &  -20.52  &  0.14 & 87   &  0.306  &  -19.00  &  0.47   \\
20  &  0.663  &  -19.58  & -1.00 & 89   &  1.151  &  -22.46  &  0.41   \\
28  &  0.247  &  -21.41  &  0.04 & 90   &  0.412  &  -19.52  &  1.00  \\
30  &  0.751  &  -21.78  &  0.34 & 91   &  0.513  &  -20.13  &  1.00   \\
32  &  1.024  &  -22.54  & -1.00 & 92   &  0.381  &  -20.94  &  0.50   \\
33  &  0.707  &  -22.07  &  1.00 & 93   &  0.377  &  -19.95  &  1.00   \\
38  &  1.208  &  -24.05  &  1.00 & 96   &  0.290  &  -22.21  &  1.00   \\
44  &  0.672  &  -21.88  &  0.51 & 100  &  0.303  &  -19.93  &  1.00   \\
45  &  0.132  &  -18.15  &  0.53 & 102  &  0.824  &  -22.39  &  0.20   \\
46  &  0.912  &  -22.04  &  0.08 & 103  &  1.159  &  -23.70  &  1.00   \\
47  &  0.173  &  -17.38  &  1.00 & 108  &  0.588  &  -21.25  & -1.00   \\
49  &  0.707  &  -20.98  &  0.37 & 111  &  0.302  &  -18.47  &  0.21   \\
50  &  0.538  &  -21.22  &  0.09 & 118  &  0.816  &  -21.36  &  0.36   \\
51  &  0.536  &  -21.11  & -1.00 & 123  &  0.095  &  -19.04  &  0.11 \\
54  &  0.418  &  -20.67  & -1.00 & 124  &  0.671  &  -20.81  & -1.00 \\
55  &  0.815  &  -21.58  & -1.00 & 125  &  0.873  &  -22.00  &  0.24   \\
56  &  0.318  &  -17.80  &  0.23 & 127  &  0.695  &  -21.39  &  1.00 \\
59  &  0.418  &  -20.58  &  0.50 & 143  &  1.102  &  -22.31  &  0.09 \\
60  &  1.392  &  -19.54  & -1.00 & 147  &  0.514  &  -21.24  &  1.00 \\
67  &  0.588  &  -19.45  & -1.00 & 148  &  0.876  &  -22.41  & -1.00 \\
69  &  0.692  &  -21.67  &  1.00 & 150  &  0.795  &  -22.11  &  1.00 \\
70  &  0.348  &  -20.62  &  0.38 & 152  &  0.617  &  -22.27  &  0.17 \\
71  &  0.132  &  -16.72  &  1.00 & 154  &  0.614  &  -22.09  &  1.00 \\
72  &  0.787  &  -21.07  &  0.07 & 155  &  0.665  &  -20.71  &  1.00 \\
73  &  0.822  &  -23.08  &  1.00 & 161  &  0.960  &  -22.65  &  1.00 \\
75  &  0.724  &  -20.13  & -1.00 & 166  &  0.378  &  -18.76  &  1.00  \\
77  &  1.020  &  -22.11  &  0.36 & 172  &  0.378  &  -19.93  & -1.00 \\
78  &  0.823  &  -21.49  &  0.55 & 204  &  0.709  &  -19.77  & -1.00   \\
\hline
\end{tabular}
\end{minipage}
\end{table}

\subsection{Completeness as function of redshift}

In order to be sure that we are studying the same kind of objects at different
redshifts we must determine the absolute limiting magnitude of our sample. On
doing this we avoid biasing our sample to brighter objects at high redshift.
Some claims of galactic evolution have been a consequence of this bias. As an
example, Simard et al (1999) analyzed the problem of the completeness of the
sample. If selection effects were ignored in their galaxies, then the mean disc
surface brightness increases by $\approx 1.3$ magnitudes from $z=0.1$ to
$z=0.9$. Most of this evolution is plausibly due to comparing low--luminosity
galaxies in nearby redshift bins to high-luminosity galaxies in distant bins.
If this effect is taken into account, no discernible evolution remains in the
disc surface brightness of their disk dominated galaxies. In order to avoid
this kind of problem it is necessary to make a selection of the objects
on the basis of their absolute luminosity.

Given our apparent limiting magnitude of $I=23$, we have studied the
completeness of our sample for three different class of galaxies: Sa, Sc and E.
Figure 3 shows, for a limiting magnitude of I=23 , the absolute magnitude down
to which a galaxy can be observed as a  function of $z$. This figure was
generated using spectral models of 15 Gyr old galaxies (Poggianti 1997). For our
distribution of 120 objects with $I\leq23$, the parameters which maximize the
number of objects into a complete sample are $z\leq0.8$ and $M_I\leq-20.0$
($M_B\leq-18$)\footnote{We have repeated this calculation with the starburst
galaxy NGC4449 without finding any substantial difference.}. This left us with a
total  of 61 objects: 20 E/S0, 25 spirals, 16 irregulars and mergers, or
equivalently:$\sim33\%$ E/S0 $\sim41\%$ spirals and $\sim26\%$ unclassified
objects. Figure 4 shows the $M-z$ diagram for our whole sample down to
$I\leq23$. The complete subsample studied (left-down corner)  is enclosed by
the horizontal and vertical lines. This kind of selection criteria is similar
to that used by Simard et al (1999) and Fried et al (2001).

\section{Results and discussion}

\subsection{Galaxy classification}

There is no substantial differences between the fraction of galaxies which
correspond to the different classes when the sample is restricted in apparent
and absolute magnitude.  We must note  that these numbers are in good agreement
with the percentage of E/S0 given by visual classification systems (van den
Bergh et al 1996, 2000) in the HDF, but are quite different from that given
automated classification of Marleau \& Simard (1998),  8\%. The classification
of Marleau \& Simard (1998) takes into account all objects with $I_{814}\le 26$
from the HDF. They claim that  the discrepancy with visual classifications is
due to the difference in the classification of small round galaxies with
half-light radii less than 0.$''$31. Visually these galaxies are classified as
elliptical galaxies, Marleau \& Simard classify them as disc--dominated
systems with bulge fractions less than 0.5. But,
galaxies with an intrinsically big $B/T$ tends to be systematically obtained
with lower values of $B/T$ in their automated routine (see Fig. 10 of Marleau
\& Simard 1998). The  ellipticals in the HDF which have been (probably)
mis--classified by these authors are those principally coming from the fainter
subsample. Although the relation between the $B/T$ output of their simulation
and the input magnitude of the objects  is not provided by these authors (and
consequently our assertions must be taken with caution), it is certainly
possible that  $B/T$ results stronger affected at increasing the input
magnitude (i.e. at lower signal ratios), and for that reason, the high redshift
population of elliptical galaxies remains biased. By using an automated
procedure which avoids this problem we have been able to obtain a result
similar  to  van den Bergh et al (1996, 2000).

Interestingly, our sample and the HDF one are imaging a galaxy population
centered around $z\sim0.5$, the principal difference being in the  exposure
time. Because of the different depth in the images, substantial differences
would be expected for the fainter subpopulation (smaller and irregular
galaxies)  between our sample and those based on the HDF. In fact, in the HDF
apparent magnitude--limited sample contains 39\% of unclassified objects
whereas we obtain $\sim30\%$. 

\subsection{Galaxy evolution}

 The two main models of galaxy evolution (monolithic collapse and hierarchical
clustering) present a completely different scenario  of galaxy evolution, so
that the observational implications  also are very different. One of  these
concerns  the comoving density of the galaxies. In the redshift interval
studied, the hierarchical model framework proposes the comoving density of big
galaxies (E/S0s and spirals) decrease with redshift, being constant in
the monolithic model. We have computed the comoving density $\rho(z)$ for these
two different types of galaxies in our complete subsample.

We have assumed both a linear function for modeling the comoving
density,

\begin{equation} 
 \rho(z)=a+bz  
 \end{equation} 
 
  with a=$\rho(0)$ the comoving density at z=0 and
b=$(\rho(z_{max})-\rho(0))/z_{max}$ with $z_{max}$ the maximum value which z
reaches in our limiting subsample, and  a power law of the form:

\begin{equation} 
 \rho(z)=a(1+z)^b
 \end{equation}

with a=$\rho(0)$ the comoving density at z=0.

To reduce the loss of information in  our data we  avoid binning them. The
values of the parameters of the function $\rho(z)$  are achieved  by running a
Kolmogorov-Smirnov test between the cumulative probability distribution
function  of finding a galaxy inside our imaging solid angle at a given z given
$\rho(z)$  and the cumulative distribution  from the real data. The cumulative
probability function for our model is computed as:

\begin{equation} 
P(z)=\frac{\int_{z_{min}}^{z}\rho(z')r^2(z')(dr/dz)dz'}
{\int_{z_{min}}^{z_{max}}\rho(z')r^2(z')(dr/dz)dz'}
\end{equation} 
where $z_{min}$ is the closest galaxy redshift, $z_{max}$=0.8 for our
limiting subsample and r(z) is the comoving distance to an object placed at z.

The Kolmogorov-Smirnov (KS) test gives the probability that two data sets come
from the same distribution. The best comoving density is that which maximize
this probability.  

\begin{table*}[!htb]
 \centering
  \caption{Parameters of the comoving density. H$_0$=75kms$^{-1}$Mpc$^{-1}$}
  \begin{tabular}{ccccccc}
\hline
Model & Type& a & b  & c  & KS p & KS p$_F$\\
\hline
 & &$\Omega_m=1$ & & $\Omega_\Lambda=0$ & \\
\hline
a+b$\times$z &E/S0&0.0033($\pm0.0015$) &-0.0015($\pm$0.0010) &  - &  0.90 & 0.87\\
a$\times$(1+z)$^b$ &E/S0&0.0039($\pm0.0018$) &-1.6($\pm$0.4) & -  &  0.92 & -\\
a+b$\times$z &S&0.0069($\pm0.0025$) &0.0014($\pm$0.0006)  &  - &  0.78 & 0.70\\
a$\times$(1+z)$^b$ &S&0.0060($\pm0.0031$) &1.7($\pm$0.5)   & - &  0.92 & -\\
a+b$\times$z+c$\times$z$^2$&S&0.0095($\pm0.0036$)&0.0027($\pm$0.0012)&-0.0031($\pm$0.0018)&0.96&-\\
\hline
 & &$\Omega_m=0.3$& & $\Omega_\Lambda=0.7$ &\\
\hline
a+b$\times$z&E/S0&0.0049($\pm0.0026$) &0.0032($\pm$0.0040) &  - &  0.86&0.83 \\
a$\times$(1+z)$^b$ &E/S0&0.0055($\pm0.0024$) &-1.1($\pm$0.3) & -& 0.82&- \\
a+b$\times$z &S&0.0071($\pm0.0022$) &-0.0019($\pm$0.0027) &  - &  0.81& 0.73\\
a$\times$(1+z)$^b$&S&0.0043($\pm0.0025$) &1.5($\pm$0.4) & - &  0.85&-\\
a+b$\times$z+c$\times$z$^2$&S&0.0063($\pm0.0042$)&0.0017($\pm$0.0011)&-0.0012($\pm$0.0015)&0.87&-\\
\hline
\end{tabular}
\end{table*}

As a matter of caution, we must note that E/S0 galaxies are placed
preferentially in high density environments, being more strongly clustered than
other types of galaxies. In order to evaluate the effect of  clustering in our
comoving density, we have studied the contribution of the E/S0 galaxies of each
field to our cumulative function. The number of galaxies on the regions of more
accumulation (0.48-0.54 and 0.69-0.73) come from both fields with approximately
the same contribution, rejecting a clustering explanation. 

To evaluate the errors on the parameters in the E/S0 and in the spiral
sample we have assumed that a least two galaxies in each sample  are
misclassified. This represents $\sim$10\% of each sample. We construct all the
subsamples that can be obtained by removing two elements on the original
samples and then we recover the values of the parameters associated to them.
Using these values we estimate the median and the standard deviation. These are
the numbers that we present as the parameter  estimations and the errors
associated to these measurements. Fig. 5 shows the original whole sample (i.e.
without removing any point) and overplotted is the cumulative function
associated with the parameters measured as explained before. Bar errors in Fig.
5 were estimated by measuring at each point  the maximum distance between the
cumulative function represented by using the whole sample and all the
cumulative functions resulting from the previous subsamples. We have also
overplotted  the cumulative distribution obtained from the comoving densities
fitted by Fried et al. (2001), who have a similar absolute magnitude cut for
their sample ($M_B\leq-18.5$). 

The comoving density of the E/S0s which gives a maximum probability in
the KS test for a linear form is given by: $\rho(z)=0.0033 (\pm0.0015) -0.0015
(\pm 0.0010)\times z$. The KS probability of this  density is 0.90. This
comoving density is closer to that deduced by Fried et al (2001). Using their
fit to our sample we obtain a KS p (KS p$_F$) of 0.87. The number of
E/S0s  decreases with redshift. For the cosmology chosen, this decrease
is  $\sim45(\pm30)\%$.  At using the power--law model, the KS probability is 
slightly better, 0.92. We have: $\rho(z)=
0.0039(\pm0.0018)\times(1+z)^{-1.6(\pm0.4)}$. In this case, the decrease of
elliptical galaxies is $\sim60(\pm10)\%$. This behavior is in a very good
agreement with the prediction from the  hierarchical clustering scenario for
this cosmology (Baugh et al. 1996; Kauffmann et al. 1996) but differs from the
results presented in Totani \& Yoshii (1998) and Im et al. (2001).
Interestingly, Daddi (2001) has pointed out that strong discrepancies in the
number density evolution for the EROs (Extremely Red Objects\footnote{Most of
these objects are expected to be E/S0s at redshift 1$\leq$z$\leq$2.}) can be
understood in terms of cosmic variance: ``it is much probable, on average, to
underestimate the true ERO surface density with small area surveys''. Maybe a
similar explanation also holds for more modest redshift E/S0 population and
this can be of  help to understand the  discrepancies in the number density
evolution pointed out for different authors.

For the spirals, the comoving density is $\rho(z)=0.0069 (\pm0.0025)+ 0.0014
(\pm 0.0006)\times z$, but the KS probability is just 0.78. The parameters of
the power--law model for this family are $\rho(0)=0.0060 (\pm0.0031)$ and
m=1.7$(\pm0.5)$ with a KS probability of 0.92. Interestingly, for this
population a peak in the range $z=0.4-0.5$ is shown in the  comoving density
obtained from  binned data in Fried et al. (2001), although they  fit only a
linear comoving density.  Probing on this possibility, we have also tested a
quadratic comoving density: 
\begin{equation}
  \rho(z)=a+bz+cz^2  
\end{equation}
where the interpretation of these parameters is as follows: a=$\rho(0)$,
b=2$\triangle\rho/z_p$ where $z_p$ is the redshift where the comoving density
reaches its biggest value and $\triangle\rho=\rho(z_p)-\rho(0)$ and
c=-$\triangle\rho/z_p^2$. Using a quadratic comoving density we obtain the
highest probability, 0.96, with the next values for the parameters:
$\rho(z)=0.0095 (\pm 0.0036) + 0.0027 (\pm 0.0012) \times z -0.0031 (\pm
0.0018) \times z^{2}$. Notice that this implies a peak of the density  at
$z=0.43$. Nevertheless, the value of the comoving density at this peak is just
6\% higher that at $z=0$. Meanwhile the value of the density at $z=0.8$ is
slightly higher (about 1\%) than at $z=0$. Consequently, contrary to the
E/S0s, brighter spiral galaxies ($M_B\leq-18$) seem to have a relatively
quiet evolution.

Our values of $\rho(0)$ for E/S0 and spiral galaxies are in good
agreement with the values that can be obtained by using the fit to the
Schechter luminosity function (Schechter 1976) of nearby samples (Marzke et al.
1998). For $M_B\leq-18$, the local comoving density is
$\rho(0)$=$0.0026\pm0.0007$ (E/S0s) and $\rho(0)$=$0.0054\pm0.0014$
(spiral galaxies).

We have also evaluated the previous quantities assuming a different cosmology:
$\Omega_m=0.3$ and  $\Omega_\Lambda=0.7$. In this case, our absolute magnitude
limit is $M_B\leq-19$. We summarize our results in Table 3. The E/S0 comoving
density at this cosmology seems to no--evolve or slightly decreases. This is a
similar result to that obtained for this cosmology by Totani \& Yoshii (1998)
and Im et al. (2001) and what it is expected from semi--analytical hierarchical
models (e.g. Kauffmann \& Charlot 1998).  The results for the spiral galaxies
are compatible with no number density evolution.

\section{Conclusions}

We present quantitative morphology of galaxies in two Hawaiian Deep Fields
imaged by 6 HST fields. Down to the limiting magnitude of our sample, nearly
all galaxies  have spectroscopic redshifts. The morphology has been obtained by
fitting a pure S\'ersic and S\'ersic + exponential profiles to the surface
brightness distribution of the galaxies. Monte-Carlo simulations have been
carried out in order to determine the limiting magnitude down to which the
recovered structural parameters  are reliable.  The galaxies have been
classified according to the $B/T$ ratio.  E/S0s systems are those with
$B/T>0.5$. Our simulations suggest an apparent magnitude limit of I=23.
We  have also accurately determined the absolute limiting magnitude of our
sample $M_B\leq-18$. The complete subsample is composed by 61 objects up to
$z=0.8$.

The percentage of the different galaxy types in the whole sample are in good
agreement with those obtained in the HDF by visual methods. We have computed
the comoving density of the galaxies as function of redshift. For an
Einstein--de Sitter universe, the comoving density of  E/S0s decreases as z
increases, in very good agreement with the predictions of hierarchical
clustering models of galaxy evolution. The comoving density of spiral galaxies
shows a good fit to a quadratic form: it grows a $\sim$6\% from $z=0$ until
$z=0.43$, and then decreases slightly  until $z=0.8$. This fit is compatible
with no number evolution. For open or $\Lambda$ universes,  the E/S0 galaxies
comoving density  is compatible with no number density evolution or a slightly
decrease as it is expected from semi--analytical models in hierarchical
clustering scenarios. Density comoving for brighter spiral galaxies also remain
quite constant at this redshift range.

\section*{Acknowledgments}

We wish to thank David Crist\'obal and Juan Betancort  for valuable
discussions. We are  also indebted to Victor Debattista who kindly proofread
versions of this manuscript. The authors are grateful to the anonymous referee
for the valuable suggestions that helped us to improve the rigor and clarity of
this paper.

Based on observations with the NASA/ESA $Hubble$
$Space$ $Telescope$ obtained at the Space Telescope Science Institute, which is
operated by the Association of Universities for Research in Astronomy, Inc.,
under NASA contract 5-26555.

\bsp

\label{lastpage}

\end{document}